\documentstyle[12pt,psfig]{article}
\makeatletter
\def\@cite#1#2{{[{#1}]\if@tempswa\typeout
{IJCGA warning: optional citation argument
ignored: `#2'} \fi}}
\newcount\@tempcntc
\def\@citex[#1]#2{\if@filesw\immediate\write\@auxout{\string\citation{#2}}\fi
  \@tempcnta\z@\@tempcntb\m@ne\def\@citea{}\@cite{\@for\@citeb:=#2\do
    {\@ifundefined
       {b@\@citeb}{\@citeo\@tempcntb\m@ne\@citea\def\@citea{,}{\bf ?}\@warning
       {Citation `\@citeb' on page \thepage \space undefined}}%
    {\setbox\z@\hbox{\global\@tempcntc0\csname b@\@citeb\endcsname\relax}%
     \ifnum\@tempcntc=\z@ \@citeo\@tempcntb\m@ne
       \@citea\def\@citea{,}\hbox{\csname b@\@citeb\endcsname}%
     \else
      \advance\@tempcntb\@ne
      \ifnum\@tempcntb=\@tempcntc
      \else\advance\@tempcntb\m@ne\@citeo
      \@tempcnta\@tempcntc\@tempcntb\@tempcntc\fi\fi}}\@citeo}{#1}}
\def\@citeo{\ifnum\@tempcnta>\@tempcntb\else\@citea\def\@citea{,}%
  \ifnum\@tempcnta=\@tempcntb\the\@tempcnta\else
   {\advance\@tempcnta\@ne\ifnum\@tempcnta=\@tempcntb \else
\def\@citea{--}\fi
    \advance\@tempcnta\m@ne\the\@tempcnta\@citea\the\@tempcntb}\fi\fi}
\makeatother

\newcommand{\lt}{\ln{m_t^2\over Q^2}}

\newcommand{\gsim}{\lower.7ex\hbox{$\;\stackrel{\textstyle>}{\sim}\;$}}
\newcommand{\lsim}{\lower.7ex\hbox{$\;\stackrel{\textstyle<}{\sim}\;$}}
\newcommand{\be}{\begin{equation}}
\newcommand{\ee}{\end{equation}}
\newcommand{\bea}{\begin{eqnarray}}
\newcommand{\eea}{\end{eqnarray}}
\newcommand{\SUSY}{\makebox[1.25cm][l]{$\line(4,1){35}$\hspace{-1.15cm}{SUSY}}}

\def\baselinestretch{1}
\begin{document}
%%%%%%%%%%%%%%%%%%%%%%%%%%% subequations.sty %%%%%%%%%%%%%%%%%%%%%%%%
\catcode`@=11
\newtoks\@stequation
\def\subequations{\refstepcounter{equation}%
\edef\@savedequation{\the\c@equation}%
  \@stequation=\expandafter{\theequation}%   %only want \theequation
  \edef\@savedtheequation{\the\@stequation}% % expanded once
  \edef\oldtheequation{\theequation}%
  \setcounter{equation}{0}%
  \def\theequation{\oldtheequation\alph{equation}}}
\def\endsubequations{\setcounter{equation}{\@savedequation}%
  \@stequation=\expandafter{\@savedtheequation}%
  \edef\theequation{\the\@stequation}\global\@ignoretrue

\noindent}
\catcode`@=12
%%%%%%%%%%%%%%%%%%%%%%%%%%%%%%%%%%%%%%%%%%%%%%%%%%%%%%%%%%%%%%%%%%%%%
\begin{titlepage}

\title{{\bf
Phenomenology of low-scale supersymmetry breaking models}} \vskip2in
\author{
{\bf Ignacio Navarro$$\footnote{\baselineskip=16pt E-mail: {\tt
ignacio.navarro@durham.ac.uk}}}
\hspace{3cm}\\
%\vskip.35in
 $$~{\small IPPP, University of Durham, DH1 3LE Durham, UK}.
}
\date{}
\maketitle
\def\baselinestretch{1.15}
\begin{abstract}
\noindent In this letter we consider the distinctive phenomenology
of supersymmetric models in which the scale of SUSY breaking is
very low, $\sqrt{F}={\cal O}(\rm TeV)$, focusing on the Higgs
sector and the process of electroweak breaking. Using an effective
Lagrangian description of the interactions between the observable
fields and the SUSY breaking sector, it is shown how the
conventional MSSM picture can be substantially modified. For
instance, the Higgs potential has non-negligible SUSY breaking
quartic couplings that can modify completely the pattern of
electroweak breaking and the Higgs spectrum with respect to that
of conventional MSSM-like models.

\end{abstract}

\thispagestyle{empty} \vspace{5cm}  \leftline{}

\vskip-20cm \rightline{} \rightline{IPPP/03/40}
\rightline{DCPT/03/92} \vskip3in

\end{titlepage}
%%%%%%%%%%%%%%%%%%%%%%%%%%%%%%%%%%%%%%%%%%%%%%%%%%%%%%%%%%%%%%%%%%%
\setcounter{footnote}{0} \setcounter{page}{1}
\newpage
\baselineskip=20pt

\section{Introduction}

\noindent

Supersymmetry (SUSY), despite all the new, recently proposed
solutions to the gauge hierarchy problem, is still (probably) the
best candidate for new physics above the electroweak scale. The
fact that it can stabilize ``tree level'' mass scales against
radiative corrections, thus solving the gauge hierarchy problem,
is its more appreciated feature, but it has many other (good)
characteristics. It is not free, however, of phenomenological and
theoretical problems (for reviews, see \cite{Haber:1984rc}).
Almost all of them are related to the fact that SUSY, if realized
in nature, has to be broken. The way this breaking takes place is
usually the more obscure point of supersymmetric models. In any
case, it is encouraging (and non-trivial) that one can build
phenomenologically viable models in which SUSY is broken, all
scalar partners of Standard Model (SM) fermions acquire a positive
mass squared and the electroweak breaking takes place in a
satisfactory way.

The usual approach in phenomenological studies of supersymmetric
models is to write down a supersymmetric Lagrangian and to
supplement it with a set of supersymmetry breaking  ($\SUSY$)
terms that parameterize the effect of the breaking, without making
any assumption about the nature of the breaking itself. The only
thing that is required is that these terms do not generate
quadratic divergences in the renormalization of scalar masses
(therefore they are called $soft$ terms). However, when doing
this, we $are$ making an assumption about the nature of the
$\SUSY$: we are assuming that the breaking takes place at a very
high energy scale. The reason for this is the following. Suppose
SUSY is broken in some sector of the theory in which there are
some fields whose F-terms take some vacuum expectation value
(v.e.v.), of order $F$. This breaking will be transmitted to every
other sector of the theory. The scalar $\SUSY$ breaking masses in
any other sector (coupled to this one only through interactions
suppressed by powers of some mass scale $M$) are typically of
order $m_s\sim F/M$. However, the appearance of $soft$ breaking
terms is not the only effect of the breaking, since the so-called
$hard$ breaking terms are also generated. For instance, $\SUSY$
quartic couplings for the scalars are typically of order
$\lambda_h \sim m_s^2/M^2$ \cite{Polonsky:2000rs,Brignole:2003cm}.
These hard breaking terms are always suppressed by inverse powers
of $M$ with respect to the soft terms, so in the case of a big
hierarchy $M\gg \sqrt{F} \gg m_s$, neglecting this terms makes a
good approximation. However, although in most models of $\SUSY$
there is in fact a hierarchy in the scales involved, we have no
real information available about what these scales are, and for
all that we know they could be of order $TeV$.

In this letter, following \cite{Brignole:2003cm}, we will review
the (quite distinctive) phenomenological properties of theories in
which all these scales are of order $TeV$, so the hard $\SUSY$
terms play a relevant role in the phenomenology. Notice that these
terms, although $do$ generate quadratic divergencies in the
renormalization of scalar masses, $do$ $not$ destabilize any
hierarchy since they are suppressed by powers of the mass scale
$M$, that is to be identified with the cut-off of the effective
description of $\SUSY$ that we are considering. Actually, the fact
that SUSY protects mass scales against radiative corrections is
not due to the absence of quadratic divergences in the quantum
corrections to the theory, since in fact there are quadratic
divergences in its quantum corrections (see \cite{Brignole:2000kg}
for compact formulae encoding all one loop radiative corrections
to a general 4D $N=1$ SUSY theory). The reason why mass scales are
stabilized against quantum corrections can be traced back to the
fact that all radiative corrections can be incorporated to the
K\"ahler potential, since we know from non-renormalization
theorems that the superpotential is not renormalized in
perturbation theory. Since the K\"ahler potential is a real
function of the fields with mass dimension two, the only
$relevant$ operator ($i.e.$ the only term in which a positive mass
dimension coupling could appear, so that quantum corrections could
drive it to be of the order of the cut-off) would be one with a
single field, that should be a singlet $\delta K \sim X S$, with
$S$ a singlet and $X$ a parameter with dimensions of mass that
would naturally be of the order of the cut-off. However, in global
SUSY such terms in the K\"ahler potential are not meaningful since
they do not give any contribution to the Lagrangian, so quantum
corrections will never destabilize mass scales in global
SUSY\footnote{This is not the case for supergravity. Once one
makes SUSY a local symmetry those terms do appear in the
Lagrangian, and it has been shown that, despite of being a
supersymmetric theory, the presence of singlets that couple with
the Higgses can destabilize the electroweak scale once one
computes the radiative corrections arising to the theory in
supergravity \cite{Bagger:1993ji}.}.

In this brief review, we will focus on the low energy
phenomenology and we will pay particular attention to the
interplay of SUSY and electroweak breaking and to the Higgs sector
spectrum in low scale $\SUSY$ models. 
New possibilities for electroweak breaking show up
\cite{Brignole:2003cm}, this breaking generically takes place in a
less fine tuned way \cite{ceh} and the lightest Higgs boson can be
much heavier than in conventional MSSM scenarios
\cite{Polonsky:2000rs,Brignole:2003cm}. Another important feature
of these scenarios is that the gravitino is superlight (with a
mass $m_{3/2}=F/(\sqrt{3}M_p$)), and its couplings with MSSM
fields are in general not suppressed, so one can find
characteristic collider signatures of such a superlight gravitino
\cite{Dicus:1989gg} (see \cite{Brignole:1998uu} for other
phenomenological implications of a superlight gravitino).

\section{Supersymmetry breaking: effective description.}

Almost all models that try to address the problem of how $\SUSY$
is transmitted to the observable sector share a common structure:
it is assumed that SUSY is broken in some ``hidden'' sector, at a
scale $F$, and that this sector couples to the MSSM fields only
trough non-renormalizable interactions suppressed by powers of
some high energy mass scale $M$. Different models propose
different physics generating these effective interactions. For
instance, in gravity mediation these are the interactions
generated by the structure of the supergravity Lagrangian, and
thus this scale is the Planck mass, $M=M_p$. Fixing $m_s\sim
F/M_p$ to be of order $TeV$ to solve the hierarchy problem we get
$\sqrt{F}$ to be ${\cal O}(10^{11}GeV)$. Another example is gauge
mediation, where it is assumed that there are some fields charged
under the SM gauge group that couple at tree level to the $\SUSY$
sector. If these fields have a large mass $M$, below this mass we
get an effective theory in which the MSSM fields are coupled to
the $\SUSY$ sector through non-renormalizable interactions
(generated at the loop level) suppressed by powers of $M$.

Generically, the specific details of the mediation mechanism
generate specific patterns of soft breaking terms. It is not a
coincidence that all these models have this common structure. The
reason for this is that if one tries to break supersymmetry in a
renormalizable model one finds that the supertrace of the mass
matrix is identically zero even when SUSY is broken. This means
that the sum of the masses of fermions is equal to the sum of the
masses of bosons. So if we try to break supersymmetry using only
the MSSM fields and a renormalizable Lagrangian we will always
find superparticles lighter than some ordinary particles. This
difficulties are overcome in models in which the transmission of
$\SUSY$ to MSSM particles can be described using an effective
non-renormalizable Lagrangian (that is valid only up to some high
energy scale $M$). In this spirit, the approach that we will
follow is to describe the transmission of $\SUSY$ using effective
interactions, without relying on any specific microscopic dynamics
that can generate it. This new physics should be now close to the
electroweak scale and it could be some massive fields (analogously
to the case of gauge mediation), or could have a more fundamental
character, as in supersymmetric models with a low scale of quantum
gravity, see for instance \cite{Casas:2001xv}. This scale of new
physics, although close to the electroweak scale, still has to be
somewhat larger than it, so it makes sense to consider the
effective theory below it, (but above the electroweak scale). We
will assume that after integrating out these heavy degrees of
freedom we are left with a globally supersymmetric theory whose
degrees of freedom are just those of the MSSM plus a singlet
chiral superfield, $T$, responsible of $\SUSY$ \footnote{We will
not bother about the problem of destabilizing divergencies arising
in supergravity theories in the presence of singlets since the
field $T$ need not be a singlet above the scale $M$.}.

Our starting point will be the $N$=1 globally supersymmetric Lagrangian

\be \label{leffgen} {\cal L} = \int \!\! d^4 \theta K (\bar{\phi},
e^{2 V} \phi) + \left[ \int \!\! d^2 \theta \, W(\phi) +  {\rm
h.c.} \right] + {1 \over 4} \left[ \int \!\! d^2 \theta
f_{ab}(\phi) {\cal W}^a {\cal W}^b +  {\rm h.c.} \right]  \ee
where $K(\bar{\phi},\phi)$, $W(\phi)$ and $f_{ab}(\phi)$ are the
effective K\"ahler potential, superpotential and gauge kinetic
functions, respectively, and higher derivative terms are
neglected. We decompose chiral superfields according to
$\phi^i \Rightarrow \phi^i + \sqrt{2} \psi^i \theta +  F^i \theta\theta 
+ \ldots$ and vector superfields
according to $V^a  \Rightarrow A^a_{\mu} 
\theta \sigma^{\mu} \bar{\theta} + 
(\lambda^a \theta \bar{\theta}\bar{\theta} + {\rm h.c.})
+ {1 \over 2} D^a \theta \theta \bar{\theta}\bar{\theta}$,
in the Wess-Zumino gauge. The effective Lagrangian for the component fields can
be obtained by a standard procedure \cite{Wess:cp} (see also the
appendix of \cite{Casas:2001xv}). In particular, the scalar
potential has the general expression
\be
\label{Vgeneral}
V=V_F + V_D = W_i K^{i\bar{\jmath}} \bar{W}_{\bar{\jmath}}
\ + \
{1\over 2} \left[K_i ({\bf t}_a\phi)^i + \xi_a \right]
f_R^{ab}
\left[K_j ({\bf t}_b\phi)^j + \xi_b \right] \, .
\ee
Subscripts denote derivatives ($W_i \equiv \partial W/\partial \phi^i$,
$\bar{W}_{\bar{\jmath}} \equiv \partial {\bar W}/
\partial \bar{\phi}^{\bar{\jmath}} \equiv (\partial W/\partial \phi^j)^*$,
$K_i \equiv \partial K/\partial \phi^i$,\ldots),
$K^{i\bar{\jmath}}$ is the inverse of the K\"ahler metric
$K_{\bar{\imath}j} \equiv \partial^2 K/\partial
\bar{\phi}^{\bar{\imath}} \partial \phi^j$ and $f_R^{ab}$ is the
inverse of the metric $(f_R)_{ab} \equiv {\rm Re} f_{ab}$  of the
vector sector ({\it i.e.} $K^{i\bar{\jmath}}K_{\bar{\jmath}\ell}
=\delta^i_{\ell}$ and $f_R^{ab} (f_R)_{bc}=\delta^a_c$). The order
parameter for supersymmetry breaking, which will be non-zero by
assumption, is \be \label{fdef} F^2 \equiv \langle V \rangle =
\langle V_F + V_D \rangle = \langle {\bar F}^{\bar{\imath}}
K_{\bar{\imath}j} F^j + {1 \over 2} D^a (f_R)_{ab} D^b \rangle \,
, \ee where the v.e.v.s of the auxiliary fields are \be
\label{auxvevs} \langle F^i \rangle = - \langle
K^{i\bar{\jmath}}\bar{W}_{\bar{\jmath}} \rangle
 \;\; , \;\;
\langle D^a \rangle = - \langle f_R^{ab} \left[K_j ({\bf
t}_b\phi)^j + \xi_b \right] \rangle \ . \ee
Fermion mass terms
have the form $ - {1 \over 2} (\lambda^a,\psi^i) {\cal M}
(\lambda^b,\psi^j)^T + {\rm h.c.}$, where the matrix ${\cal M}$ is
given by
\be
\label{generalM}
\hspace{-0.5cm} {\cal M} = \left(
\begin{array}{cc}
- \displaystyle{1\over 2} \langle (f_{ab})_{\ell} F^{\ell} \rangle
&
\sqrt{2} \langle K_{\bar{\ell} j} (\overline{{\bf t}_a \phi})^{\bar{\ell}}
+ \displaystyle{1\over 4}  (f_{ac})_j D^c \rangle
\\ &  \\
 \sqrt{2} \langle K_{\bar{\ell} i} (\overline{{\bf t}_b \phi})^{\bar{\ell}}
+ \displaystyle{1\over 4}  (f_{bc})_i D^c \rangle
&
\langle W_{ij} +  {\bar F}^{\bar{\ell}} K_{\bar{\ell}ij} \rangle
\end{array}
\right)\ .
\ee
By using the extremum conditions of the scalar potential and gauge
invariance, it is easy to check that the mass matrix $ {\cal M}$
has an eigenvector $( {1\over \sqrt{2}} \langle D^b \rangle ,
\langle F^j \rangle)^T$ with zero eigenvalue, which corresponds to
the goldstino state. This eigenvector specifies the components of
goldstino field $\tilde{G}$ contained in the original fields
$\psi^i$ and  $\lambda^a$. We also recall that, in the framework
of local SUSY, the goldstino degrees of freedom become the
longitudinal components of the gravitino, which obtains a mass
$m_{3/2}=F/(\sqrt{3}M_p)$. When $\sqrt{F}$ is close to the
electroweak scale, $m_{3/2}$ is much smaller than typical
experimental energies, which implies that the dominant gravitino
components are precisely the goldstino ones \cite{Fayet:vd}.

The chiral superfields will be denoted by
$\phi^i=(\phi^\alpha,T)$, where $\phi^{\alpha}$ are the MSSM
chiral superfields (containing Higgses, leptons and quarks) and
$T$ is the singlet superfield whose auxiliary field v.e.v.
$\langle F^T \rangle$ breaks SUSY. For small fluctuations of the
fields $\phi^{\alpha}$, the expansions of $K$, $W$ and $f_{ab}$
can be written as \bea \label{keff} K & = & k(\bar{T},T) +
c_{\bar{\alpha} \beta} (\bar{T},T)
\bar{\phi}^{\bar{\alpha}}\phi^\beta + {1 \over 2} \left[
d_{\alpha\beta}(\bar{T},T) \phi^\alpha \phi^\beta + {\rm
h.c.}\right] + \ldots
\\
\label{weff}
W & = &  w(T) + {1\over 2} \mu_{\alpha\beta}(T)
\phi^\alpha \phi^\beta  + {1 \over 3!} h_{\alpha\beta\gamma}(T)
\phi^\alpha \phi^\beta \phi^\gamma + \ldots
\\
\label{feff}
f_{ab} & = & f_a(T) \delta_{ab} + \ldots
\eea
The functions $c_{\bar{\alpha} \beta}, d_{\alpha\beta},
h_{\alpha\beta\gamma}, f_a$ are assumed to depend on $T$ through
the ratio $T/M$, where $M$ is the typical scale of suppression of
the non-renormalizable operators of the theory. This theory is
therefore an effective description of some yet more fundamental
theory, and is valid only up to this scale, $M$. Then the induced
SUSY-breaking mass splittings within the $\phi^{\alpha}$ and $V^a$
multiplets are characterized by a scale $\tilde{m}\sim F/M$. We
also make the standard assumption that $\mu_{\alpha\beta}$, if
non-vanishing, has size ${\cal O}(\tilde{m})$ rather than ${\cal
O}(M)$, {\it i.e.}~$\mu_{\alpha\beta}(T) \sim (F/M)
{\tilde\mu}_{\alpha\beta}(T/M)$.

As we said, standard scenarios are characterized by a strong
hierarchy $M \gg \sqrt{F} \gg \tilde{m}$. In this limit the
physical components of the $T$ multiplet ({\it i.e.}~the goldstino
and its scalar partners, the `sgoldstinos') are almost decoupled
from the other fields, and the effective theory for the
$\phi^{\alpha}$ and $V^a$ multiplets is well approximated by a
renormalizable one. The latter is characterized by gauge couplings
$g_a^2 = 1/\langle {\rm Re}f_a \rangle_0$, an effective
superpotential $\hat W$ and a set of soft SUSY breaking terms,
whose mass parameters are ${\cal O}(\tilde{m})$. This is the usual
MSSM scenario. The MSSM parameters can be computed in terms of the
functions appearing in $K$, $W$ and $f_{ab}$ above. Let us
consider for simplicity the case of diagonal matter metric, {\it
i.e.}~$c_{\bar{\alpha} \beta} = c_{\alpha} \delta_{\bar{\alpha}
\beta}$, and rescale the fields in order to have canonical
normalization: $\langle \sqrt{c_\alpha} \rangle_0 \phi^\alpha
\rightarrow \phi^\alpha$, $\langle \sqrt{ {\rm Re}f_a } \rangle_0
V^a  \rightarrow V^a$. The effective superpotential of the
renormalizable theory is \be \hat{W} = {1\over 2}
\hat{\mu}_{\alpha\beta} \phi^\alpha \phi^\beta + {1 \over 3!}
\hat{h}_{\alpha\beta\gamma} \phi^\alpha \phi^\beta \phi^\gamma\ ,
\ee where \bea \label{mupar} \hat{\mu}_{\alpha\beta} & = &
\left\langle \frac{ \mu_{\alpha\beta} + \bar{F}^{\bar{T}}
\partial_{\bar{T}} d_{\alpha\beta} } { (c_\alpha c_\beta)^{1/2} }
\right\rangle_0 \ ,
\\
\hat{h}_{\alpha\beta\gamma} & = &  \left\langle
\frac{ h_{\alpha\beta\gamma} }{ (c_\alpha c_\beta c_\gamma)^{1/2} }
\right\rangle_0 \ .
\eea
Soft breaking terms are described by
\be
{\cal L}_{\rm soft} =  - \tilde{m}_\alpha^2 |\phi^\alpha|^2
- \left[ {1\over 2} (\hat{\mu}_{\alpha\beta} B_{\alpha\beta})
\phi^\alpha \phi^\beta
+ {1 \over 3!} (\hat{h}_{\alpha\beta\gamma} A_{\alpha\beta\gamma})
\phi^\alpha \phi^\beta \phi^\gamma +
 {1\over 2} M_a \lambda^a \lambda^a + {\rm h.c.} \right],
\ee
where
\bea
\label{softmass}
\tilde{m}_\alpha^2 & = &  \left\langle - |F^T|^2
\partial_T \partial_{\bar{T}}
\log c_\alpha \right\rangle_0 \ ,
\\
B_{\alpha\beta} & = & \left\langle - F^T  \partial_T
\log \left( \frac { \mu_{\alpha\beta} + \bar{F}^{\bar{T}}
\partial_{\bar{T}} d_{\alpha\beta} }
{c_\alpha c_\beta} \right) \right\rangle_0 \ ,
\\
A_{\alpha\beta\gamma} & = & \left\langle - F^T  \partial_T
\log \left( \frac
{ h_{\alpha\beta\gamma} } {c_\alpha c_\beta c_\gamma} \right)
\right\rangle_0 \ ,
\\
\label{mgaugi} M_a & = & \left\langle - F^T  \partial_T \log (
{\rm Re} f_a) \right\rangle_0 \ . \eea The formulae presented
above can be obtained also taking a specific limit ($M_p
\rightarrow \infty$, $m_{3/2} \rightarrow 0$ with $F=
\sqrt{3}m_{3/2} M_p$ fixed) of supergravity results \cite{SUGRA}.

In the case of strong hierarchy of scales these are all the
$\SUSY$ effects one needs to consider at low energies. The $T$
multiplet has plays an external role in the derivation: it only
provides the SUSY breaking v.e.v. $\langle F^T \rangle$. However,
if the scales $M$ and $F$ are not much larger than the TeV scale
and the ratio $F/M^2 \sim \tilde{m}/M \sim \tilde{m}^2/F$ is not
negligible, the standard MSSM picture is corrected by additional
effects and novel features emerge. The components of MSSM
multiplets ($\phi^\alpha$ and $V^a$) have novel non-negligible
interactions among themselves as well as non-negligible
interactions with the physical components of $T$ (goldstino and
sgoldstinos). Moreover, since some of the $\phi^\alpha$ fields
({\it i.e.}~the Higgses) have to obtain a v.e.v. in order to break
the gauge symmetry, in principle one should reconsider the
minimization of the scalar potential taking into account both $T$
and such fields. In addition, the $F$ components of the Higgs
multiplets and the $D$ components of the neutral vector multiplets
could give non-negligible contributions to SUSY breaking. In this
case the goldstino could have components along all neutral
fermions ($\tilde T$, Higgsinos and gauginos).

%%%%%%%%%%%%%%%%%%%%%%%%figure%%%%%%%%%%%%%%%%%%%%%%%%
\begin{figure}[t]
%\psdraft
\centerline{
\psfig{figure=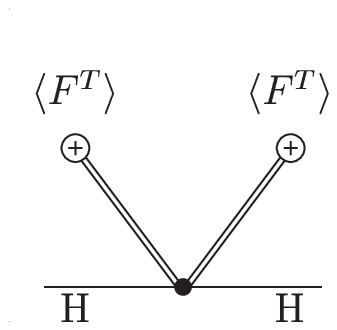,height=4cm,width=6cm,angle=0}} \caption
{\footnotesize Diagramatic origin of the soft mass for the
Higgses.} \label{diag1}
\end{figure}
%%%%%%%%%%%%%%%%%%%%%%%%%figure%%%%%%%%%%%%%%%%%%%%%%%%

To see the origin of these hard breaking terms consider for
instance a coupling $\delta K =-{\alpha \over M^2} |T|^2 |H|^2$ in
the K\"ahler potential. This term will give a contribution to the
soft mass of the field $H$ (once the F-term of $T$ takes a v.e.v.)
given by $\alpha m_s^2$ where $m_s^2 \equiv \langle |F^T|^2
\rangle / M^2$ as can be seen from eq.(\ref{softmass}). One can
visualize this contribution as coming from the diagram depicted in
Fig. 1, where double lines represent auxiliary fields and crossed
circles represent the v.e.v. $\langle F^T \rangle_0$. However,
this term will also give contributions to a quartic coupling for
$H$ coming from the diagram depicted in Fig. 2. So the total
$\SUSY$ contribution to the potential of $H$ coming from such a
term in the K\"ahler potential will be \be \delta {\cal L} =\alpha
\; m_s^2 |H|^2 + {2 \alpha^2 m_s^2 \over M^2} |H|^4 + \dots \ee
\noindent where the dots represent higher order non-renormalizable
terms. As for the case of the soft terms one can compute and give
analytic expressions for all the $\SUSY$ terms that will be
generated in the Lagrangian for the general theory with arbitrary
$K$, $W$ and $f_{ab}$ that we are considering. We will focus now
on the Higgs sector and discuss how these terms can modify
significantly the pattern of electroweak breaking.

%%%%%%%%%%%%%%%%%%%%%%%%figure%%%%%%%%%%%%%%%%%%%%%%%%
\begin{figure}[t]
%\psdraft
\centerline{
\psfig{figure=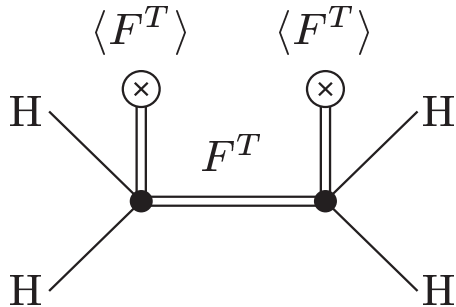,height=4cm,width=6cm,angle=0}} \caption
{\footnotesize Diagramatic origin of quartic SUSY breaking
couplings for the Higgses.} \label{diag1}
\end{figure}
%%%%%%%%%%%%%%%%%%%%%%%%%figure%%%%%%%%%%%%%%%%%%%%%%%%

\section{Electroweak breaking in low-scale supersymmetry breaking models.}

When $F/M^2$ is not negligible, some higher order terms in the
expansions of $K$, $W$ and $f_{ab}$ not written explicitly in
eqs.~(\ref{keff}), (\ref{weff}) and (\ref{feff}) can become
important. In order to find all the possible renormalizable terms
that can appear in the Higgs potential we will have to consider
all the ${\cal O}(H^4)$ terms in $W$ and $K$. As anticipated in
the previous section, the coefficient functions appearing in $W$
and $K$ will depend on the field $T$ and on some mass scales. Thus
we write: \bea W & = & w(T) + \mu(T) H_1\cdot H_2 + {1 \over 2
M}\ell(T)\left(H_1 \cdot H_2\right)^2 + \ldots \label{W}
\\
K & = &
k(\bar{T},T) + c_1(\bar{T},T)|H_1|^2  + c_2(\bar{T},T)|H_2|^2
+ \left[ d(\bar{T},T) H_1\cdot H_2 + {\rm h.c.}\right]
\nonumber
\\
& + & {1\over 2M^2} e_1(\bar{T},T)|H_1|^4
+{1\over 2M^2} e_2(\bar{T},T)|H_2|^4
+{1\over M^2}e_3(\bar{T},T)|H_1|^2 |H_2|^2
\nonumber
\\
& + & {1\over M^2}e_4(\bar{T},T)|H_1\cdot H_2|^2
+\left[ {1 \over 2M^2} e_5(\bar{T},T) (H_1\cdot H_2)^2\right.
\nonumber
\\
&+&\left. {1\over M^2}e_6(\bar{T},T)|H_1|^2 H_1\cdot H_2 +{1\over
M^2}e_7(\bar{T},T)|H_2|^2 H_1\cdot H_2 +  {\rm h.c.} \right]+
\ldots \label{Kahler} \eea The K\"ahler potential $K$ is assumed
to contain a single mass scale $M$. Thus the coefficient functions
$c_i$, $d$ and $e_i$ in $K$ are in fact dimensionless functions of
$T/M$ and $\bar{T}/M$ while $k(\bar{T},T) \sim M^2 \tilde{k}
(\bar{T}/M,T/M)$. On the other hand, $W$ should contain, besides
$M$, the SUSY-breaking scale $F$ (notice that $F\sim \langle
\partial_T W \rangle$). Although in our effective field theory approach it is not possible to determine
what is the precise dependence on $M$ and $F$ of the coefficient
functions in $W$, a reasonable criterion is to insure that each
parameter of the component Lagrangian in the $T$-$H_1$-$H_2$
sector receives contributions of the same order from  $K$ and $W$.
An example of this are the two contributions to the effective
$\hat{\mu}$ parameter in Eq.~(\ref{mupar}).  The plausibility of
this criterion is  stressed by the fact that there is a
considerable freedom to move terms between $K$ and $W$ through
analytical redefinitions of the superfields. Consequently, we can
assume
\be
\label{wscal}
w(T) \sim F M \tilde{w}(T/M) \ , \;\;
\mu(T) \sim {F \over M} \tilde{\mu}(T/M) \ , \;\;
\ell(T) \sim {F \over M^2} \tilde{\ell}(T/M) \ ,
\ee
where $ \tilde{w},\tilde{\mu},\tilde{\ell}$ are dimensionless
functions of their arguments. For the expansion of the gauge
kinetic functions $f_{ab}$ it is enough for our purpose to keep
${\cal O}(H^2)$ terms. The indices in $f_{ab}$ are saturated with
those of the super-field-strengths ${\cal W}^a {\cal W}^b$, see
eq.~(\ref{leffgen}), so the allowed irreducible representations in
$f_{ab}$ are those contained in the symmetric product of two
adjoints. For the $SU(2)\times U(1)$ gauge group, such
representations are singlet, triplet and fiveplet: \be f_{ab} =
f^{(s)}_{ab}+ f^{(t)}_{ab}+ f^{(f)}_{ab} \ee The expansion of the
singlet part $f^{(s)}_{ab}$ reads \be \label{fsin}
f^{(s)}_{ab}=\delta_{ab}\left[f_a(T)+{1 \over M^2} h_a(T) \,
H_1\cdot H_2 + \ldots \right] \, . \ee The triplet part
$f^{(t)}_{ab}$ is associated with the $SU(2)$-$U(1)_Y$ cross-term
${\cal W}^A {\cal W}^Y$, where $A$ is an $SU(2)$ index, so the
non-vanishing components of $f^{(t)}_{ab}$ are
$f^{(t)}_{AY}=f^{(t)}_{YA}$ and their expansion starts at ${\cal
O}(H^2)$: \be \label{ftri} f^{(t)}_{AY} = {1 \over M^2}
\omega(T)\, ( H_1\cdot \sigma^A H_2 ) + \ldots \ee while we can
neglect the fiveplet part since its leading term is ${\cal
O}(H^4)$.

Now, from the expansions of $W$, $K$ and $f_{ab}$ that we have
considered we can compute the component Lagrangian with all
relevant renormalizable terms, and in particular the scalar
potential, which is given by the general expression in
eq.~(\ref{Vgeneral}). General formulae can be found in
\cite{Brignole:2003cm}, but for the discussion that follows we
will only use the fact that all possible terms of a general
two-Higgs-doublet model (2HDM) are generated (see
\cite{Gunion:2002zf} and references therein for a recent analysis
of 2HDMs):
\bea
\label{VTH}
V & = &
V_0(\bar{T},T) + m_1^2(\bar{T},T)|H_1|^2  + m_2^2(\bar{T},T)|H_2|^2
+ \left[ m_3^2(\bar{T},T) H_1\cdot H_2 + {\rm h.c.}\right]
\nonumber
\\
& + &
{1\over 2} \lambda_1(\bar{T},T)|H_1|^4
+{1\over 2} \lambda_2(\bar{T},T)|H_2|^4
+\lambda_3(\bar{T},T)|H_1|^2 |H_2|^2
+\lambda_4(\bar{T},T)|H_1\cdot H_2|^2
\nonumber
\\
& + &
\left[ {1 \over 2} \lambda_5(\bar{T},T) (H_1\cdot H_2)^2
+\lambda_6(\bar{T},T)|H_1|^2 H_1\cdot H_2
+\lambda_7(\bar{T},T)|H_2|^2 H_1\cdot H_2
+  {\rm h.c.} \right]\nonumber\\
&+&\ldots
\eea
It is important to keep in mind that the parametric dependence of
the coefficients in $V$ is $m_i^2 \sim {\cal O}(F^2/M^2)$ and
$\lambda_i \sim  {\cal O}(F^2/M^4) +  {\cal O}(g^2) \sim {\cal
O}(m_i^2/M^2) +  {\cal O}(g^2)$. In general, for a given
potential, one can try to perform either an exact minimization or
at least an iterative one, relying on the expansion of the
potential in powers of $H_i/M$ and on the consistent assumption
that the Higgs v.e.v.s are smaller than $M$. In the iterative approach, the starting point for the
determination of the v.e.v.s are the zero-th order values of
$\langle H_i^0 \rangle_0$ and $\langle T \rangle_0$, where
$\langle H_i^0 \rangle_0 =0$ and $\langle T \rangle_0$ is the
minimum of $V_0(\bar{T},T)$.

The form of the Higgs potential in eq.~(\ref{VTH}) already allows
us to make some general observations on the possible patterns of
electroweak breaking. Let us set $m_i^2\equiv \langle
m_i^2(\bar{T},T) \rangle_0$ for brevity. There are two necessary
conditions for electroweak breaking in a general 2HDM: the origin
of field space for the Higgses must be unstable and there must not
be unbounded from below (UFB) directions.
%%%%%%%%%%%%%%%%%%%%%%%%figure%%%%%%%%%%%%%%%%%%%%%%%%
\begin{figure}[t]
%\psdraft
\centerline{
\psfig{figure=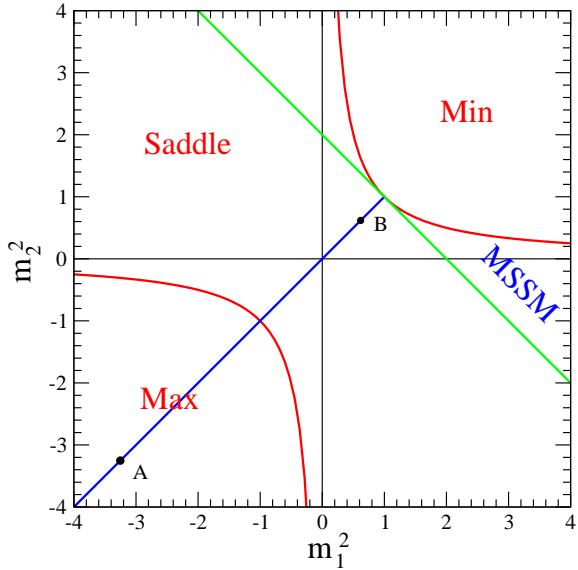,height=8cm,width=8cm,angle=-90,bbllx=3.cm,%
bblly=2.cm,bburx=20.cm,bbury=19.cm}} \caption {\footnotesize
Depending on the values of $m_1^2$ and $m_2^2$ the origin of the
Higgs potential can be destabilized or not. Absence of UFB
directions impose additional constraints on the Higgs masses for
the MSSM.} \label{thdmbreak}
\end{figure}
%%%%%%%%%%%%%%%%%%%%%%%%%figure%%%%%%%%%%%%%%%%%%%%%%%%
The first condition depends on the quadratic part of the Higgs
potential. This is a minimum, a saddle point or a maximum,
depending on the mass parameters $m_i^2$: \bea m_1^2 m_2^2 -
|m_3^2|^2 &>& 0\ ,\ m_1^2+ m_2^2 \ >\ 0\ ,\hspace{1.8cm}
[\mathrm{Minimum}] \label{minimum}
\\
m_1^2 m_2^2 - |m_3^2|^2 &<& 0\ ,
\hspace{4.2cm} [\mathrm{Saddle\  Point}]
\label{saddle}
\\  m_1^2 m_2^2 - |m_3^2|^2 &>& 0\ ,\ m_1^2+ m_2^2 \ <\  0\
.\hspace{1.8cm} [\mathrm{Maximum}]
\label{maximum}
\eea
These equations define three regions in the
$\{m_1^2,m_2^2\}$-plane, labelled by  `Min', `Saddle' and `Max' in
Fig.~1 (where the axes are given in units of $|m_3^2|$). Such
regions are separated by the upper and lower branches of the
hyperbola $m_1^2 m_2^2 -|m_3^2|^2 =0$. Electroweak breaking can
take place in the regions `Saddle' or `Max', while the region
`Min' is excluded. The second condition is the absence of UFB
directions, along which the quartic part of the potential gets
destabilized. In the MSSM the quartic couplings receive only
contributions from D-terms, namely $\lambda_{1,2}={1\over 4}(g^2 +
g_Y^2)$, $\lambda_3={1\over 4}(g^2 - g_Y^2)$,  $\lambda_4=-{1\over
2}g^2$, $\lambda_{5,6,7}=0$. Then the potential  is indeed
stabilized by the quartic terms, except along the D-flat
directions $|H_1|=|H_2|$. It is then required that the quadratic
part of $V$ be positive along these directions:
\bea m_1^2 + m_2^2 - 2 |m_3^2| > 0\ .\hspace{2cm} [\mathrm{Potential\
 bounded\ from\ below}]
\label{UFB}
\eea
This condition applies {\em only} to the MSSM and
corresponds to the region of Fig.~1 above the straight line tangent
to the upper branch of the hyperbola. Since eq.~(\ref{UFB})
is incompatible  with eq.~(\ref{maximum}),
it follows that the MSSM conditions for electroweak  breaking
are given by eqs.(\ref{saddle},\ref{UFB}), as is well known.
In Fig.~1 the corresponding region is a subset of the region
`Saddle' and is labelled by `MSSM':
it is made of the (two) areas between the upper branch of
the hyperbola and the tangent line.

In the case that matters to us, when SUSY is broken at a
moderately low scale, the $\lambda_i$ couplings in (\ref{VTH})
also receive sizeable ${\cal O}(F^2/M^4)$ contributions, besides
the ${\cal O}(g^2)$ ones. Therefore condition (\ref{UFB}) is no
longer mandatory to avoid UFB directions, since the boundedness of
the potential can be ensured by imposing appropriate conditions on
the $\lambda_i$ parameters. Thus the presence of the latter
parameters extends the parameter space, relaxes the constraints on
the quadratic part of the potential and opens a lot of new
possibilities for electroweak breaking. In particular, both
alternatives (\ref{saddle}, \ref{maximum}) are now possible. This
means that most of the $\{m_1^2, m_2^2\}$ plane can in principle
be explored: only the region `Min' is excluded. For instance, now
the universal case $m_1^2= m_2^2$ is allowed. Actually, in the
MSSM these mass parameters could be degenerate at high energy and
reach non-degenerate values radiatively by RG running (falling in
the region `MSSM' of Fig.~1, typically with $m_1^2>0$, $m_2^2<0$).
The fact that $m_2^2$ is the only scalar mass that tends to get
negative in this process is considered one of the virtues of the
MSSM, in the sense that $SU(2)\times U(1)$ breaking is
``natural''. Now, we see that even if the universal condition
holds at low-energy we can still break $SU(2)\times U(1)$.

As opposed to the radiative breaking, now electroweak breaking
generically occurs already at tree-level. Still, it is ``natural''
in a sense similar to the MSSM. For example, if all the scalar
masses are positive and universal, $SU(2)_L\times U(1)_Y$  is the
only symmetry that can be broken because (with R-parity conserved)
the only off-diagonal bilinear coupling among MSSM fields is
$m_3^2 \, H_1\cdot H_2$, which can drive symmetry breaking in the
Higgs sector if condition (\ref{saddle}) is satisfied. Finally,
the fact that quartic couplings are very different from those of
the MSSM changes dramatically the Higgs spectrum and properties
(which will be tested at colliders, see {\it e.g.}
\cite{Carena:2002es}). In particular, as will be clear from the
example in the next subsection, the MSSM bound on the lightest
Higgs field does no longer apply. Likewise, the fact that these
couplings can be larger than the MSSM ones reduces the amount of
tuning necessary to get the proper Higgs v.e.v.s \cite{ceh}.

\section{A simple example.}

In this section we present, for illustrative purposes, a simple
example (example "A" of \cite{Brignole:2003cm}) where many of the
previously discussed unconventional features are clear. For
simplicity we consider a model that is symmetric under the
interchange $H_1$ and $H_2$, so we can use the general formulae
for symmetric potentials of the appendix of \cite{Brignole:2003cm}
to obtain the minimization conditions. Despite this symmetry, the
model can accommodate both $\tan\beta=1$ and $\tan\beta \neq 1$
(depending on the choice of parameters). The superpotential, gauge
kinetic functions and K\"ahler potential are chosen as \be W =
\Lambda_S^2 T + \mu H_1\cdot H_2 + {\ell\over 2M}(H_1\cdot H_2)^2
\ , \;\;\;\; f_{ab} = {\delta_{ab}\over g_a^2} \left(1 + 2 {\eta_a
\over M} T \right)\ , \ee and \bea K & = & |T|^2 +  |H_1|^2 +
|H_2|^2 \nonumber\\ & - & {\alpha_t \over 4 M^2} |T|^4 + {\alpha_1
\over M^2}|T|^2 \left(|H_1|^2+|H_2|^2\right) + {e_1 \over
2M^2}\left(|H_1|^4+|H_2|^4\right)\ ,  \eea where all parameters
are taken to be real, with $\alpha_t>0$. We will sometimes use the
auxiliary parameter ${\tilde m}=\Lambda_S^2/M$.

We will analyze the model perturbatively in the Higgs v.e.v.s. We
will only retain the first terms of the expansion, which will be
sufficient to illustrate the main qualitative features of this
example. At zero-th order, {\it i.e.} for vanishing Higgs v.e.v.s,
we have $\langle T\rangle_0=0$, SUSY is broken by  $\langle F^T
\rangle_0 = -\Lambda_S^2$, $\tilde{T}$ is the goldstino and the
complex $T$ field has mass $m_T^2 = \alpha_t\tilde{m}^2$. The
effects of electroweak breaking start to appear at next order,
{\it i.e.} when the potential $V(T,H_1,H_2)$ is minimized and the
Higgses take v.e.v.s. In particular, since $V(T,H_1,H_2)$ contains
${\cal O}(T H^2)$ cubic terms, $T$ receives a small induced v.e.v.
$\langle T\rangle=\alpha_1\mu v_1v_2/(\alpha_t\Lambda_S^2)$ and
$T$-$H$ mass mixing appears. Instead of keeping the field $T$
together with the Higgses, however, we find it more convenient to
use the alternative method of decoupling it. We integrate out $T$
and study a reduced effective potential for the Higgs doublets
only. This choice is also supported by the special fact that all
Higgs boson masses turn out to be ${\cal O}(\lambda v^2)$ in this
model, {\it i.e.} naturally lighter than the $T$ mass, which is
${\cal O}({\tilde m}^2)$.

The Higgs v.e.v.s and spectrum are determined by an effective
quartic potential $V(H_1,H_2)$ with particular values for its mass
terms: \be m_1^2=m_2^2=\mu^2-\alpha_1 \tilde{m}^2\ ,\;\;\;\;
m_3^2=0\ , \ee and quartic couplings \bea
\lambda_1=\lambda_2&=&{1\over
4}(g^2+g_Y^2)+2\alpha_1^2{\tilde{m}^2\over M^2}\ ,\nonumber\\
\lambda_3&=&{1\over 4}(g^2-g_Y^2)+{2\over
M^2}(\alpha_1^2\tilde{m}^2-e_1\mu^2)\ ,\nonumber\\
\lambda_4&=&-{1\over 2}g^2-2\left(e_1+2{\alpha_1^2\over
\alpha_t}\right){\mu^2\over M^2}\
,\nonumber\\  \lambda_5&=&0\ ,\nonumber\\
\lambda_6=\lambda_7&=&{\ell\mu\over M}\ . \eea Applying the
general formulae given in the appendix of \cite{Brignole:2003cm}
to write down the minimization conditions that give $v^2$ and
$\sin 2\beta$ in terms of the parameters of the potential we see
that concerning the value of $\tan\beta$, we have two possible
solutions   \be |\tan\beta| = 1 \ , \ee and \be
\sin2\beta={\ell\mu/M\over ({g^2+g_Y^2})/4+2\hat{e}_1\mu^2/M^2}\ ,
\label{tbneq1} \ee where we use $\hat{e}_1\equiv
e_1+\alpha_1^2/\alpha_t$.  Both solutions are possible depending
on the choice of parameters, and in both cases ${\rm
sgn}(\tan\beta)= -{\rm sgn}(\ell\mu/M)$. It is not restrictive to
take $\ell\mu/M <0$, so that $\tan\beta >0$. Using this
convention, the explicit expressions for the Higgs masses are the
following. \be
\begin{array}{ll}
\hspace{2.5cm}{\underline{\tan\beta=1}}:&
\hspace{2.5cm}{\underline{\tan\beta\neq 1}}:\\ &\\
m_h^2={\displaystyle 2\left(\alpha_1^2{\tilde{m}^2\over
M^2}-\hat{e}_1{\mu^2\over M^2} +{\ell\mu\over M}\right)v^2}\ , &
m_h^2={\displaystyle \left[{1\over
4}(g^2+g_Y^2)+2\alpha_1^2{\tilde{m}^2\over M^2}+{\ell\mu\over
M}s_{2\beta}\right]v^2}\ ,\vspace{0.2cm}\\
m_H^2={\displaystyle\left[{1\over 4}(g^2+g_Y^2)+2\hat{e}_1{\mu^2\over
M^2} -{\ell\mu\over M}\right]v^2}\ ,&m_H^2={\displaystyle  -\left[{1\over
4}(g^2+g_Y^2)+2\hat{e}_1{\mu^2\over M^2}\right]v^2c^2_{2\beta}}\
,\vspace{0.2cm}\\ m_A^2={\displaystyle -{\ell\mu\over M} v^2 }\ ,&
m_A^2={\displaystyle  -\left[{1\over
4}(g^2+g_Y^2)+2\hat{e}_1{\mu^2\over M^2}\right]v^2}\
,\vspace{0.2cm}\\    m_{H^\pm}^2={\displaystyle \left[{1\over
4}g^2+(2\hat{e}_1-e_1){\mu^2\over M^2} -{\ell\mu\over M}\right]v^2}\ , &
m_{H^\pm}^2={\displaystyle  -\left({1\over 4}g_Y^2+e_1{\mu^2\over
M^2}\right)v^2}\ .
\end{array}
\ee
Notice that acceptable solutions with
$\tan\beta=1$
can be obtained even if we set $e_1=0$, which further
simplifies the model. To obtain solutions with $\tan\beta \neq 1 $,
however, we need $e_1 < 0$. Also notice that, in the phase
with  $\tan\beta \neq 1 $, the value of $\tan\beta$ is only
determined up to an inversion ($\tan\beta \leftrightarrow 1/\tan\beta$),
which in fact leaves the spectrum invariant. This is a consequence
of the original discrete symmetry, and we can conventionally
take $\tan\beta \geq 1$.

%%%%%%%%%%%%%%%%%%%%%%%%figure%%%%%%%%%%%%%%%%%%%%%%%%
\begin{figure}[t]
%\psdraft
\centerline{
\psfig{figure=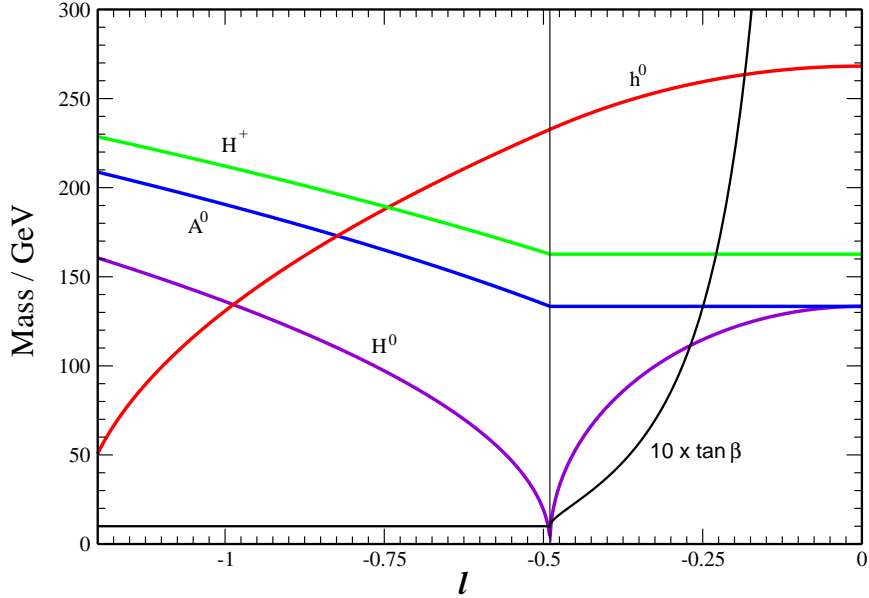,height=8cm,width=8cm,angle=-90,bbllx=3.cm,%
bblly=5.cm,bburx=20.cm,bbury=22.cm}} \caption {\footnotesize Higgs
spectrum as a function of the superpotential parameter $\ell$.
Also shown is $\tan\beta$ (scaled by a factor 10).} \label{especA}
\end{figure}
%%%%%%%%%%%%%%%%%%%%%%%%%figure%%%%%%%%%%%%%%%%%%%%%%%%

In Fig.~4 we show a numerical example where both phases of the
model are visible. We have fixed $\mu/M=0.6$, $e_1=-1.3$,
$\tilde{m}/M=0.5$, $\alpha_t=3.0$, $\alpha_1\simeq
\mu^2/\tilde{m}^2+\epsilon^2$ (with $0<\epsilon^2\ll 1$) and vary
$\ell$. For each parameter choice, the overall mass scale $M$ is
adjusted so as to get the right value of $v=246$ GeV. The closer
$\alpha_1$ is to $\mu^2/\tilde{m}^2$ the larger  $M/v$ can be. The
figure shows the Higgs spectrum and the parameter $\tan\beta$
(scaled by a factor 10 for clarity) as a function of the coupling
$\ell$. For $\ell\leq \ell_0$, with \be \ell_0\equiv {M\over
\mu}\left[{1\over 4}(g^2+g_Y^2)+2\hat{e}_1{\mu^2\over M^2}
\right]\ , \ee  the minimum lies at $\tan\beta=1$, while for $\ell
> \ell_0 $, $\tan\beta$ increases with $\ell$. For the choice of
parameters used in this figure, $\ell_0\simeq -0.49$. The spectrum
is continuous across the critical value $\ell_0$, although the
mass of the `transverse' Higgs, $H^0$, goes through zero, as was
to be expected on general grounds for symmetric potentials. We see
that, except in the neighborhood of $\ell_0\simeq -0.49$ or for
too negative values of $\ell$, the Higgs masses are sufficiently
large to escape all current experimental bounds (which are also
lower than usual due to singlet admixture, although this is
typically a small effect). There is even a region of parameters
for which $h^0$  is the heavier of the Higgses, beyond the usual
limit of $m_{h^0}\lt 200$ GeV \cite{EQ} that applies in generic
SUSY models only when they are perturbative up to the GUT scale.

\section{Conclusions}

When the scales of $\SUSY$ ($\sqrt{F}$) and $\SUSY$ mediation
($M$) are close to the electroweak scale, the usual MSSM, where
the effects of SUSY breaking in the observable sector are encoded
in a set of soft SUSY-breaking terms of size ${\cal O}(F/M)$, may
not give an accurate enough description of electroweak scale
physics. Additional effects can be relevant, in particular
non-negligible contributions to hard breaking terms, such as
${\cal O}(F^2/M^4)$ contributions to quartic Higgs couplings. In
fact, the latter contributions can compete with (and may take
over) the usual ($D$-term induced) MSSM quartic Higgs couplings,
giving rise to a quite unconventional Higgs sector phenomenology.

In this brief review of the phenomenology of these models we have
focussed on the Higgs sector, and we have analyzed, following \cite{Brignole:2003cm}, the interplay
of SUSY and electroweak breaking. We have seen how the Higgs
potential resembles that of a 2HDM, where the quadratic and
quartic couplings can be traced back to the original couplings in
the effective superpotential and K\"ahler potential. The goldstino
supermultiplet can have non-negligible interactions with the Higgs
fields, and its scalar component can also mix with them as a
result of electroweak breaking.

The presence of extra quartic couplings that may be larger than
the usual ones opens novel opportunities for electroweak breaking.
The breaking process is effectively triggered at tree-level and
presents important differences with the usual radiative mechanism.
Electroweak breaking can occur in a much wider region of parameter
space, {\it i.e.} for values of the low-energy mass parameters
that are normally forbidden. A further advantage of the extra
quartic couplings is that their presence reduces the amount of
tuning necessary to get the correct Higgs v.e.v.s. These new
quartic couplings also imply that the spectrum of the Higgs sector
is dramatically changed, and the usual MSSM mass relations are
easily violated. In particular, the new quartic couplings allow
the lightest Higgs field to be much heavier ($\lsim 500$ GeV) than
in usual supersymmetric scenarios.

We have illustrated these facts by means of an example, in which
we have analyzed the Higgs potential and the electroweak breaking
process, making clear the unconventional features that emerge.

\section*{Acknowledgments}

I would like to thank A. Brignole, J.A. Casas and J.R. Espinosa for the many things I learned collaborating with them in this subject.

%%%%%%%%%%%%%%%%%%%%%%%%%%%%%%%%%%%%%%%%%%%%%%%%%%%%%%%%%%%%%%%%%%%


\begin{thebibliography}{99}
%\cite{Haber:1984rc}
\bibitem{Haber:1984rc}
H.~E.~Haber and G.~L.~Kane,
%``The Search For Supersymmetry: Probing Physics Beyond The Standard Model,''
Phys.\ Rept.\  {\bf 117} (1985) 75;
%%CITATION = PRPLC,117,75;%%
H.~P.~Nilles,
%``Supersymmetry, Supergravity And Particle Physics,''
Phys.\ Rept.\  {\bf 110} (1984) 1;
%%CITATION = PRPLC,110,1;%%
N.~Polonsky,
%``Supersymmetry: Structure and phenomena. Extensions of the standard model,''
Lect.\ Notes Phys.\  {\bf M68} (2001) 1 [arXiv:hep-ph/0108236].
%%CITATION = HEP-PH 0108236;%%

%\cite{Polonsky:2000rs}
\bibitem{Polonsky:2000rs}
N.~Polonsky and S.~Su,
%``More corrections to the Higgs mass in supersymmetry,''
Phys.\ Lett.\ B {\bf 508}, 103 (2001) [arXiv:hep-ph/0010113];
%%CITATION = HEP-PH 0010113;%%
N.~Polonsky,
%``The scale of supersymmetry breaking as a free parameter,''
Nucl.\ Phys.\ Proc.\ Suppl.\  {\bf 101}, 357 (2001)
[arXiv:hep-ph/0102196].
%%CITATION = HEP-PH 0102196;%%

%\cite{Brignole:2003cm}
\bibitem{Brignole:2003cm}
A.~Brignole, J.~A.~Casas, J.~R.~Espinosa and I.~Navarro,
%``Low-scale supersymmetry breaking: Effective description, electroweak  breaking and phenomenology,''
Nucl.\ Phys.\ B (in press) [arXiv:hep-ph/0301121].
%%CITATION = HEP-PH 0301121;%%

%\cite{Brignole:2000kg}
\bibitem{Brignole:2000kg}
A.~Brignole,
%``One-loop Kaehler potential in non-renormalizable theories,''
Nucl.\ Phys.\ B {\bf 579} (2000) 101 [arXiv:hep-th/0001121].
%%CITATION = HEP-TH 0001121;%%

%\cite{Bagger:1993ji}
\bibitem{Bagger:1993ji}
J.~Bagger and E.~Poppitz,
%``Destabilizing divergences in supergravity coupled supersymmetric theories,''
Phys.\ Rev.\ Lett.\  {\bf 71} (1993) 2380 [arXiv:hep-ph/9307317];
%%CITATION = HEP-PH 9307317;%%
J.~Bagger, E.~Poppitz and L.~Randall,
%``Destabilizing divergences in supergravity theories at two loops,''
Nucl.\ Phys.\ B {\bf 455} (1995) 59 [arXiv:hep-ph/9505244].
%%CITATION = HEP-PH 9505244;%%

\bibitem{ceh}
J.~A.~Casas, J.~R.~Espinosa and I.~Hidalgo, to appear.

%\cite{Dicus:1989gg}
\bibitem{Dicus:1989gg}
D.~A.~Dicus, S.~Nandi and J.~Woodside,
%``Collider Signals Of A Superlight Gravitino,''
Phys.\ Rev.\ D {\bf 41} (1990) 2347;
%%CITATION = PHRVA,D41,2347;%%
%\cite{Brignole:1996fn}
A.~Brignole, F.~Feruglio and F.~Zwirner,
%``Aspects of spontaneously broken N = 1 global supersymmetry in the  presence of gauge interactions,''
Nucl.\ Phys.\ B {\bf 501} (1997) 332 [arXiv:hep-ph/9703286],
%%CITATION = HEP-PH 9703286;%%
%``On the effective interactions of a light gravitino with matter  fermions,''
JHEP {\bf 9711} (1997) 001 [arXiv:hep-th/9709111], A.~Brignole,
F.~Feruglio and F.~Zwirner,
%``Signals of a superlight gravitino at e+ e- colliders when the other  superparticles are heavy,''
Nucl.\ Phys.\ B {\bf 516} (1998) 13 [Erratum-ibid.\ B {\bf 555}
(1999) 653] [arXiv:hep-ph/9711516]; A.~Brignole, F.~Feruglio,
M.~L.~Mangano and F.~Zwirner,
%``Signals of a superlight gravitino at hadron colliders when the other  superparticles are heavy,''
Nucl.\ Phys.\ B {\bf 526} (1998) 136 [Erratum-ibid.\ B {\bf 582}
(2000) 759] [arXiv:hep-ph/9801329].
%%CITATION = HEP-PH 9801329;%%

%\cite{Brignole:1998uu}
\bibitem{Brignole:1998uu}
A.~Brignole, F.~Feruglio and F.~Zwirner,
%``Four-fermion interactions and sgoldstino masses in models with a  superlight gravitino,''
Phys.\ Lett.\ B {\bf 438} (1998) 89 [arXiv:hep-ph/9805282];
%%CITATION = HEP-PH 9805282;%%
A.~Brignole, E.~Perazzi and F.~Zwirner,
%``On the muon anomalous magnetic moment in models with a superlight  gravitino,''
JHEP {\bf 9909}, 002 (1999) [arXiv:hep-ph/9904367];
%%CITATION = HEP-PH 9904367;%%
F.~Mori and A.~A.~Natale,
%``Constraining Models With Superlight Gravitino, Scalar And Pseudoscalar Particles,''
Mod.\ Phys.\ Lett.\ A {\bf 15} (2000) 1099.
%%CITATION = MPLAE,A15,1099;%%


%\cite{Casas:2001xv}
\bibitem{Casas:2001xv}
J.~A.~Casas, J.~R.~Espinosa and I.~Navarro,
%``Unconventional low-energy SUSY from warped geometry,''
Nucl.\ Phys.\ B {\bf 620} (2002) 195 [arXiv:hep-ph/0109127].
%%CITATION = HEP-PH 0109127;%%

%\cite{Wess:cp}
\bibitem{Wess:cp}
J.~Wess and J.~Bagger,
%``Supersymmetry And Supergravity,''
{\it  Princeton, USA: Univ. Pr. (1992) 259 p}.

%\cite{Fayet:vd}
\bibitem{Fayet:vd}
P.~Fayet,
%``Mixing Between Gravitational And Weak Interactions Through The Massive Gravitino,''
Phys.\ Lett.\ B {\bf 70} (1977) 461;
%%CITATION = PHLTA,B70,461;%%
R.~Casalbuoni, S.~De Curtis, D.~Dominici, F.~Feruglio and
R.~Gatto,
%``A Gravitino - Goldstino High-Energy Equivalence Theorem,''
Phys.\ Lett.\ B {\bf 215} (1988) 313,
%%CITATION = PHLTA,B215,313;%%
%``High-Energy Equivalence Theorem In Spontaneously Broken Supergravity,''
Phys.\ Rev.\ D {\bf 39} (1989) 2281.
%%CITATION = PHRVA,D39,2281;%%

\bibitem{SUGRA}
%\cite{Hall:iz}
L.~J.~Hall, J.~Lykken and S.~Weinberg,
%``Supergravity As The Messenger Of Supersymmetry Breaking,''
Phys.\ Rev.\ D {\bf 27} (1983) 2359;
%%CITATION = PHRVA,D27,2359;%%
%\cite{Soni:1983rm}
S.~K.~Soni and H.~A.~Weldon,
%``Analysis Of The Supersymmetry Breaking Induced By N=1 Supergravity
%Theories,''
Phys.\ Lett.\ B {\bf 126} (1983) 215;
%%CITATION = PHLTA,B126,215;%%
%\cite{Giudice:1988yz}
G.~F.~Giudice and A.~Masiero,
%``A Natural Solution To The Mu Problem In Supergravity Theories,''
Phys.\ Lett.\ B {\bf 206} (1988) 480;
%%CITATION = PHLTA,B206,480;%%
%\cite{Kaplunovsky:1993rd}
V.~S.~Kaplunovsky and J.~Louis,
%``Model independent analysis of soft terms in effective supergravity and
%in string theory,''
Phys.\ Lett.\ B {\bf 306} (1993) 269 [hep-th/9303040];
%%CITATION = HEP-TH 9303040;%%
%\cite{Brignole:1997dp}
A.~Brignole, L.~E.~Ibanez and C.~Munoz
%``Soft supersymmetry-breaking terms from supergravity and
%superstring  models,''
[hep-ph/9707209].
%%CITATION = HEP-PH 9707209;%%

%\cite{Gunion:2002zf}
\bibitem{Gunion:2002zf}
F.~Boudjema and A.~Semenov,
%``Measurements of the SUSY Higgs self-couplings and the reconstruction of  the Higgs potential,''
Phys.\ Rev.\ D {\bf 66} (2002) 095007 [arXiv:hep-ph/0201219];
%%CITATION = HEP-PH 0201219;%%
J.~F.~Gunion and H.~E.~Haber,
%``The CP-conserving two-Higgs-doublet model: The approach to the  decoupling limit,''
Phys.\ Rev.\ D {\bf 67} (2003) 075019 [arXiv:hep-ph/0207010].
%%CITATION = HEP-PH 0207010;%%

%\cite{Carena:2002es}
\bibitem{Carena:2002es}
M.~Carena and H.~E.~Haber,
%``Higgs boson theory and phenomenology. ((V)),''
Prog.\ Part.\ Nucl.\ Phys.\  {\bf 50} (2003) 63
[arXiv:hep-ph/0208209].
%%CITATION = HEP-PH 0208209;%%

%\cite{Espinosa:1998re}
\bibitem{EQ}
J.~R.~Espinosa and M.~Quiros,
%``Gauge unification and the supersymmetric light Higgs mass,''
Phys.\ Rev.\ Lett.\  {\bf 81} (1998) 516 [hep-ph/9804235].
%%CITATION = HEP-PH 9804235;%%







\end{thebibliography}
\end{document}